# Absolute reflectance of a concave mirror used for astro-particle physics experiments

Razmik Mirzoyan[a], Cornelia Arcaro[b], Hanna Kellermann[c], Markus Garczarczyk[d]

a-Max-Planck-Institute for Physics, Föhringer Ring 6, D-80805 Munich, Germany, razmik.mirzoyan@mpp.mpg.de
b- Università di Padova and INFN, Via Marzolo 8, I-35131 Padova, Italy, now at Centre for Space Research, North-West University, Potchefstroom, South Africa and INAF - Osservatorio Astronomico di Padova, vicolo dell'Osservatorio 5, I-35122 Padova, Italy, cornelia.arcaro@nwu.ac.za
c- Max-Planck-Institute for Extraterrestrial Physics, Gießenbachstraße, D-85741 Garching, Germany, hkellerm@mpe.mpg.de
d-Deutsches Elektronen-Synchrotron (DESY) Zeuthen, Platanenallee 6, D-15738 Zeuthen, Germany, markus.garczarczyk@desy.de

## I. Introduction

Often the reflectivity of a concave mirror used in astro-particle physics experiments is not unambiguously defined. Sometimes the dependence of the reflectivity of the mirror surface material on the wavelength is measured with some commercial spectrophotometer-based instrument or method and it is optimistically assumed that the reflected light is entirely focused into a small spot in the focal plane of the mirror. Until recently, no proper attention was paid to the diffuse scattered light component from mirrors.

A mirror used in an astro-particle physics experiment needs to be inexpensive and just "sufficiently good" for the given purpose. As a rule, the tessellated reflectors of ground-based air Cherenkov and air fluorescence telescopes are used to collect light on the imaging cameras providing a field of view of 3.5° - 10°, i.e., their off-axis performance is of prime importance. A single mirror tile of an angular resolution of 1 - 2 arcminutes can provide a satisfactory point spread function (PSF) for a composite reflector in the range of ~3 arcminutes. This is in contrast with the mirrors used in optical astronomy where light is typically collected in the center of the camera, within a very small solid angle, and about two orders of magnitude better angular resolution is required.

The reflectivity of the surface material of a given mirror is only one of the two main parameters describing its light collection efficiency. The second parameter describes the concentration of light in its focal plane.

The radius of curvature and its variation across the surface of a mirror, together with the surface quality (roughness), define its focusing characteristics. A mirror with high surface roughness will strongly scatter light. Similarly, and obviously, if the mirror features local deviations from the specified radius of curvature across its surface, it can only poorly concentrate light into a small spot.

Pure aluminum (*Al*) and its alloys are the most-frequently-used materials for producing high-reflectivity mirrors for air Cherenkov telescopes, which are used for ground-based very high energy (VHE) gamma-ray astrophysics and air fluorescence telescopes, used for measuring cosmic air showers in the energy range $\geq 10^{17}$ eV. Cherenkov light from extended air showers arrives on the ground in the wavelength range ~(300 - 700) nm and the fluorescence light is in the near UV range of ~(300 - 400) nm. As a rule, these telescopes are set in desert locations and/or at high-mountain altitudes. Because of cost reasons, typically they have no protecting domes. Their mirrors are exposed to harsh outdoor environment. For longevity these mirrors are given a protective layer. Otherwise the *Al* would rapidly oxidize, causing a significant reflectivity loss within a time-scale of several months. The thickness of the protective layer and its variation across the mirror surface has strong impact on the reflectivity. Typically, the thickness of the protective layer of a given refractive index *n* is chosen as a multiple of λ/4, for maximizing the reflectivity at the selected optimal wavelength λ. Due to light interference, reflected from the protective and *Al* surface layers, variation of the thickness of the protective layer across the mirror makes the reflectivity not only wavelength but also position dependent. The absolute reflectance of the reflector of a telescope is the main parameter for measuring a light flux. The evaluation and monitoring of this parameter is not easy to perform for large, tessellated telescopes, operated in open-field conditions.

The measured flux and spectrum of gamma-rays from a celestial object depend inversely on the absolute reflectance of the reflector of an Imaging Air Cherenkov Telescope (IACT), making its regular monitoring crucial.

The two 17 m diameter MAGIC-I and MAGIC-II telescopes are IACTs for VHE gamma-ray astrophysics [1]. These telescopes are located at the *Roque de los Muchachos* European North Observatory on the Canary island of La Palma at the mountain altitude of 2200 m a.s.l. [2]. The prime interest of this work was to measure and regularly monitor the absolute reflectance of the reflectors. Both telescopes follow a tessellated parabolic design, with a focal length of 17 m. The imaging cameras in their foci are composed of 1039 bialkali photo multiplier-based pixels of 0.10° in diameter, providing an angular aperture of 3.5°.

Initially, the reflector of the MAGIC-I telescope was made of 964 square shaped mirrors of 495 x 495 mm² size. The reflector consisted of 11 rings of mirrors with radii of curvature between 34.0 - 36.6 m for best fitting the parabolic curvature. The focal length shortening effect for off-axis mirrors, see [3], was taken into account in the design, production and layout of the mirrors.

Every four mirrors of preselected similar focal lengths are mounted on a 60 mm thick rigid base panel made of *Al* honeycomb and are adjusted to mimic a single mirror of 1 m² area [4]. The panels are attached via the Active Mirror Control (AMC) system to the reflector frame of the telescope [5]. Each panel is fixed on three points. Two points are equipped with custom-made linear actuators, having two and three degrees of movement, respectively. The third point is a passive universal joint.

A dedicated CCD camera, mounted in the center of the reflector dish, is used for measuring the mirror spot positions in the focal plane. For this purpose, we built a remotely controlled pop-up target that can be placed in the focal plane of the telescope, just in front of the imaging camera. We installed square-shape Spectralon targets of 400 mm side length and 7 mm thickness in both imaging cameras of the MAGIC telescopes [6]. The position of the target is selected to provide sharp images of starlight from infinity.

All 964 mirrors of the MAGIC-I telescope were produced by using the so-called diamond milling technology applied to a solid 6 mm thick AlSiMg alloy. The latter was glued to a 2 cm thick *Al* honeycomb plate for providing stiffness. The rear side of the honeycomb plate was glued to a 1 mm thick *Al* plate. The sides of these mirrors were covered by epoxy glue for sealing against humidity and dust. The front side of the mirrors was given a protective quartz ($SiO_2$) layer in a successive production step [4].

The optical design of the MAGIC-II telescope is similar to that of MAGIC-I. In MAGIC-II two types of mirrors are used, both having a size of 985 x 985 $mm^2$; 143 mirrors of an *Al* alloy (AlSiMg-1%) that were diamond milled and quartz coated, and 104 mirrors of aluminized and quartz-coated glass sheets [7]. The latter are based on a 2 cm thick *Al* honeycomb that is sandwiched between two 1.7 mm thick sheets of glass, realized via the so-called cold-slumping technology that was developed by the Italian INAF in cooperation with the company Media Lario Industries [8].

The PSF has been simulated in the design report of the MAGIC telescope [9] for different incident angles of light between 0° - 1.5° as well as independently several times later by different researchers in the collaboration, showing consistent results (e.g., see [10]). The simulated on-axis PSF is about 0.04°. Individual mirror misalignment on the order of (1 - 2) arcminutes enlarges the PSF to ~0.05°.

It is important to recall that a proper IACT design should provide a focused spot that can provide efficient gamma-hadron separation, see [3] for details. The chosen optics will not smear significantly the existing size differences between the genuine gamma and hadron images. As an example, the *width* of gamma and hadron showers at TeV energies are respectively $\leq 0.15°$ and $\leq 0.3°$ [11]. To efficiently select the twice narrower and rare gamma events from the several orders of magnitude higher background of hadrons, the PSF of the reflector should not significantly exceed the size of the smallest searched for elements, i.e., that of gammas. Thus, it is natural to assume that the PSF of the reflector should stay below 0.15° in the above-mentioned case.

It is important to stress the necessity to provide this optical resolution not only in the camera center but anywhere on the camera surface area where the gamma-ray images impinge, and a strong gamma-hadron separation is desired.

To summarize, only the knowledge of the two important quantities, namely of the focusing in terms of PSF and of the surface reflectivity, allows the fully characterization of the quality of the mirrors designed for an astro-particle physics instrument.

In this report, we present the results from the application of an "in-situ" method described in [12] for measuring the absolute reflectance of a large reflector. The merits of this method are its absolute character and simplicity; the PSF and the absolute reflectance can be measured simultaneously. Below we report on our detailed studies of several parameters that play an important role for assessing the quality and the performance of a reflector. Compared to the initial report [12], these novel studies, especially those related to the scattered light component of the mirrors, helped us to seriously refine the measurement of the absolute reflectance, significantly improving its reliability and the precision that can be achieved. Furthermore, we report on our detailed comparison of the used mirror types, their advantages and drawbacks. We believe that this report could help our colleagues from the astro-particle physics community in designing and monitoring the optical performance of their telescopes.

In addition, we include in this report some measurements on the reflectivity degradation. Furthermore, we present some results on reflection losses due to scattering effects and variations of the spectral reflectivity. We performed these for individual mirrors under laboratory conditions. With these, we aim to explain the difference between the reflectivity measured locally on the surface of individual mirrors and the absolute reflectance of the MAGIC telescopes measured in their focal planes. In the end we briefly summarize our conclusions.

## II. Mirror surface reflectivity measurements

A convenient method for measuring the given mirror surface reflectivity point-wise on small areas is to use a calibrated reflectometer. We have used a reflectometer of type *IRIS 908RS* [13] to measure *in situ* the specular reflectivity component of each mirror on the MAGIC-I telescope at the wavelength of $(470 \pm 30)$ nm at two points: one at the edge and one in the center. The reproducibility of the surface reflectivity measurement was checked by taking 30 consecutive measurements from the same location. We measured a systematic error of 0.2%. The area illuminated by the device had a diameter of somewhat less than 10 mm. The measurement was performed in December 2008.

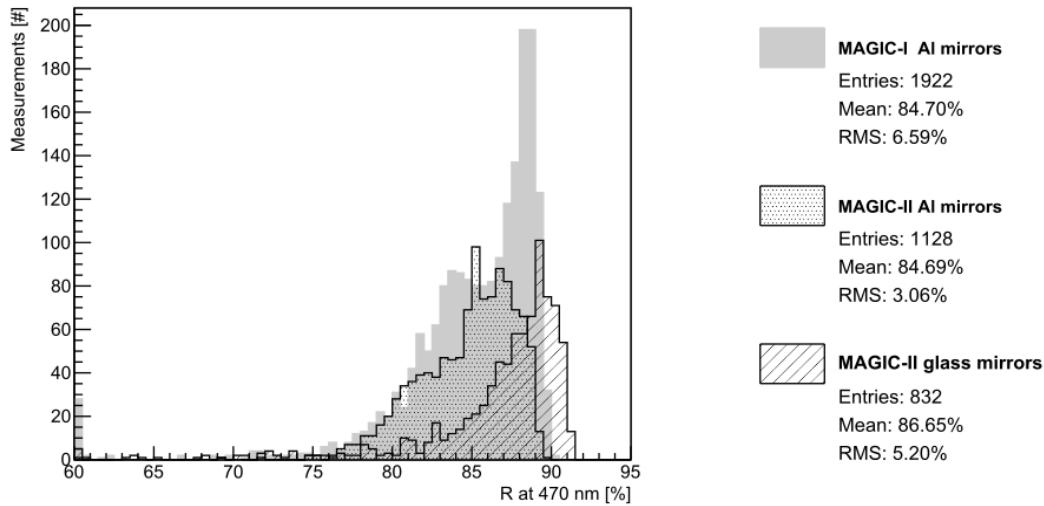

**Figure 1:** Distributions of the measured surface reflectivity of the MAGIC-I and MAGIC-II mirrors performed in December 2008 and April 2011, respectively. For MAGIC-II the different mirror types (all-*Al* and glass) are shown separately. The first bin includes the sum of measurements below 60%.

This procedure was repeated for MAGIC-II in April 2011. The average surface reflectivity of the MAGIC-I all-*Al* mirrors turned out to be (84.70 ± 6.59)%. In MAGIC-II the average reflectivity of the 143 all-*Al* mirrors was (84.69 ± 3.06)% (Figure 1).

It should be noted that the MAGIC-I mirrors were installed six years earlier than those of MAGIC-II. Apart from this, half of the mirrors, i.e., those located in the upper half of the telescope dish, were slightly scratched during the installation, resulting in ~5.2% lower reflectivity (see Figure 2 left). The study showed that the surface reflectivity degradation of all-*Al* mirrors is very low, which will be discussed in the text below. The surface reflectivity of the 104 glass mirrors of MAGIC-II was (86.65 ± 5.20)%, thus slightly higher than for the all-*Al* mirrors (see Figure 2 right). The stated errors correspond to the RMS of the measured reflectivity distributions[1].

## 1. Long-term stability of the mirror surface reflectivity

There are two sets of mirror surface reflectivity measurements available for MAGIC-I and MAGIC-II. Both were performed with the same instrument following the same measurement procedure. The time difference between the two measurements was 28 months for the MAGIC-I mirrors and 18 months for the MAGIC-II mirrors, respectively. By comparing the measured reflectivity of each individual mirror at different times, the possible reflectivity degradation was evaluated.

For the MAGIC-I mirrors, no statistically significant difference w.r.t. the averaged reflectivity of (84.70 ± 6.59)% was found within the precision of the systematic errors of these measurements. For the MAGIC-II telescope mirrors we could not find any significant surface reflectivity degradation. Note that the measurements were performed shortly after rain, which washed away the dust from the mirrors.

---

[1] We explicitly quote the RMS values, since different RMS values w.r.t. to the distribution of the two mirror types, which have been exposed for the same period to environmental impacts, are a proxy regarding the long-term behavior of the individual mirror designs.

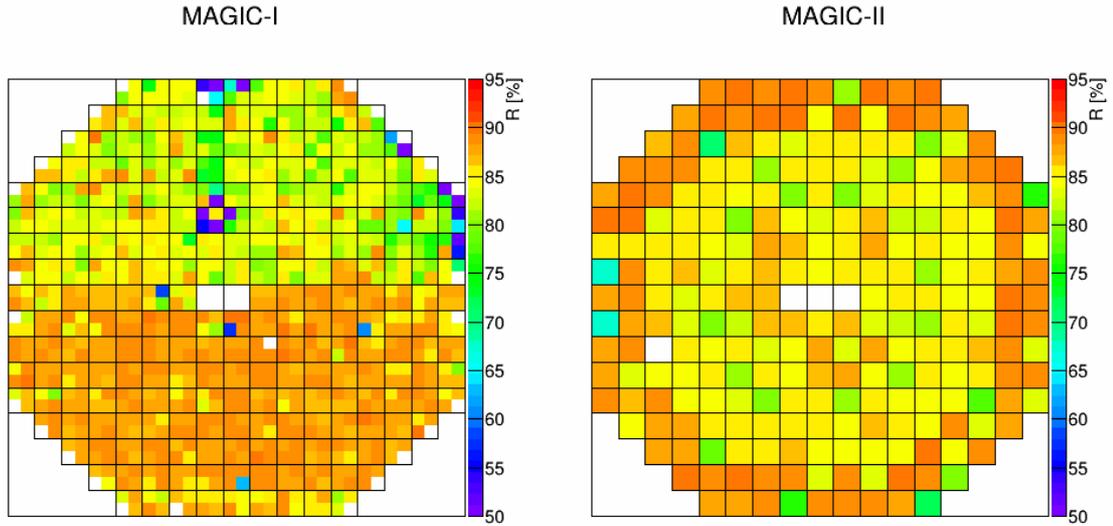

**Figure 2:** Distribution of the MAGIC-I (left) and MAGIC-II (right) mirrors on the telescope dish, front view. In MAGIC-II the all-*Al* mirrors and the glass mirrors were installed in the central and outer area of the mirror dish, respectively.

## 2. Measurements of the all-aluminum type mirror scattering

Typically, a diamond-milled all-*Al* mirror has surface imperfections at a micro-roughness level of ~(10 - 13) nm RMS [4]. During the milling process the tip of the diamond cutter removes material and deposits it on both sides of the milling path, creating humps. These elevations of material have irregular shape. Under the microscope the diamond-milled mirror looks similar to a grating from a physics laboratory, with "scratches" between the humps. In this way, some part of the impinging light will be diffusely scattered off the humps, while the other part may produce diffraction patterns somewhere in the mirror's focal plane. Also, dust deposited on the mirror surface will diffusely scatter light.

In surface reflectivity measurements such effects may remain unnoticed, when using a compact-handheld spectrograph that integrates reflected light within some non-negligible solid angle. As a result, such measurement will produce an artificially higher surface reflectivity value. However, studying the light collection efficiency, i.e., the absolute reflectance in the focus of a concave mirror, a clearly reduced reflectivity will be measured.

Below we will have a closer look into the level of the diffusely scattered losses. We assume reflection losses $R_{diff.-hump}$ for a given all-*Al* mirror by the above mentioned humps to be proportional to the ratio of the diffusely scattering area $A_{diff.-hump}$ of these humps to the total mirror area $A_{total}$:

$$R_{diff.-hump} \sim \frac{A_{diff.-hump}}{A_{total}} \quad (1)$$

This proportionality holds also for other types of surface losses as due to dust particles, micro-scratches on the mirror. We checked under the microscope randomly selected small areas on the surface of several diamond-milled all-*Al* mirrors produced by incremental cutting step size of approximately 50 μm. These showed humps of a typical width of 3 μm (see Figure 3)[2]. According to Equation (1), these correspond to a diffuse loss of 6%.

---

[2] When we obtained the first diamond milled mirrors back in 1997, we saw color disks of several cm in diameter around the focal spot of those mirrors. We reported these to the manufacturer of the diamond mirrors LT Ultra [14]. After that they started varying the step size of subsequent cuttings and the concentrated visual

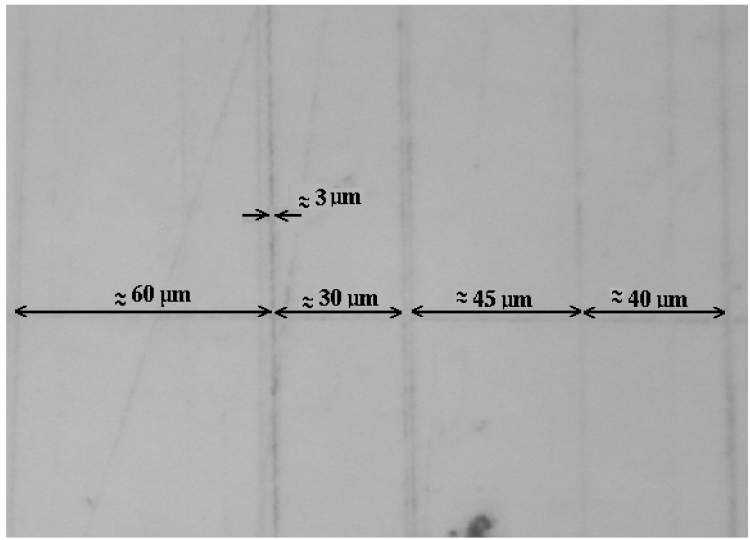

**Figure 3:** Zoomed image of a small fraction of an all-*Al* mirror seen under the microscope to examine the width of the grooves and the humps.

For estimating the reflection loss due to dust particles and scratches, we developed a dedicated image analysis: we illuminated a small region on the mirror under study and recorded its reflected image by using a CMOS camera, which was attached to a microscope. The image was then converted into black and white (two bits). The white regions correspond to areas that reflect specular light, while the black areas are due to diffuse reflection (see Figure 4). The ratio of the black area to the white provides a measure of losses. With such image analysis, reflection losses due to dust particles and scratches of up to 2-4% have been measured for the used mirrors.

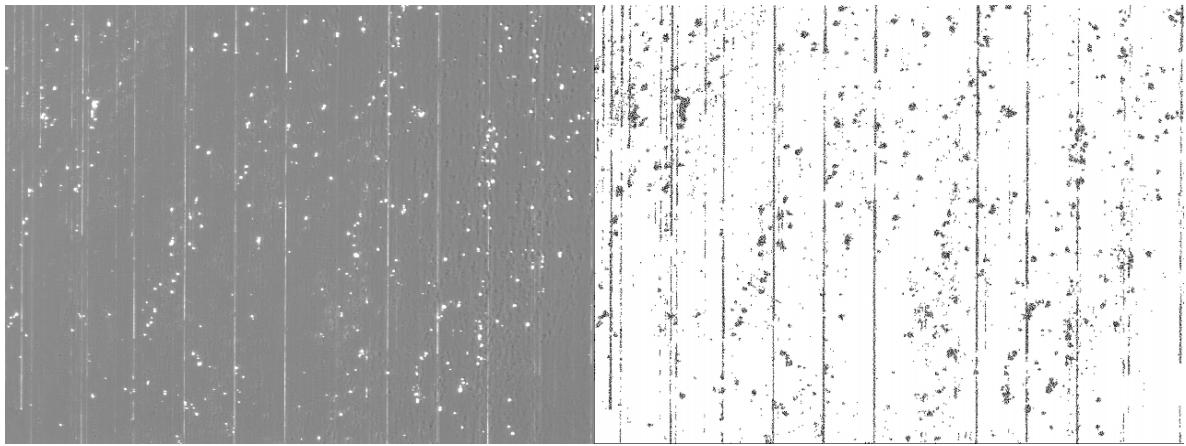

**Figure 4:** *Left:* View of a diamond-milled all-*Al* mirror surface under a microscope. Humps (white lines) on the sides of wide grooves, dust particles and scratches causing diffuse reflection can be seen.
*Right:* The same image converted into black and white (two bit).

In addition, we expected reflected light loss in the mirror focus due to diffraction effects. Depending on how regular the periods of the grooves are, the diffraction pattern is characterized by either small, exactly contoured or extended, blurred maxima (see Figure 5).

---

effect disappeared. Later on, we observed interference patterns with somewhat smeared maxima, which in our opinion witness that the manufacturer was intentionally varying the cutting step size.

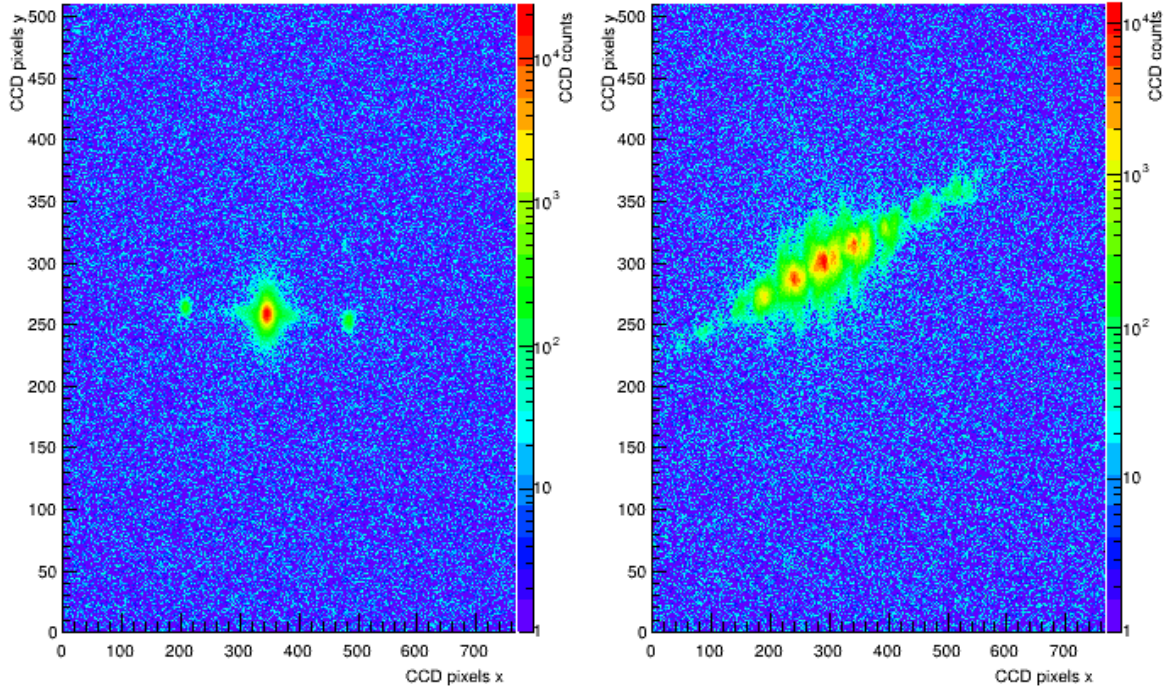

**Figure 5:** Diffraction pattern observed at a distance of approximately 3 m from two all-*Al* mirrors with constant (left) and varying (right) cutting step size.

The intensity distribution of the individual diffraction maxima was measured using a CCD camera when illuminating the mirror with a laser beam of ~1 cm diameter. Depending on the position on the mirror surface where the laser beam was incident, the first maxima orders contained up to 5% of the reflected light.

From the theory point of view, the expected reflection loss due to the total integrated scatter (*TIS;* [15]) can be estimated by the following equation:

$$TIS = \left(\frac{4\pi \cdot \sigma}{\lambda}\right)^2 \tag{2}$$

where $\sigma$ denotes the surface RMS roughness and $\lambda$ the wavelength. Assuming $\lambda = 400$ nm and $\sigma \approx (10 - 13)$ nm, measured in our studies, one obtains a scatter loss of ~(10 - 17)% correspondingly. By comparing this integral number with the sum of one by one measured losses due to the light scattered off the humps, dust, scratches and due to the diffraction, one obtains a total of ~(13 - 15)%. We conclude that these numbers are in reasonable agreement with those obtained by using Equation (2).

### 3. Study of the protective quartz layers for both mirror types

The thickness of the quartz layer for the MAGIC mirrors is chosen to optimize the reflectivity for the wavelength range where the maximum amount of Cherenkov light is detected. To study the reflectivity dependence on the quartz coating thickness, the spectral reflectivity was measured on the mirror surface (see [16] for further details) for both the MAGIC-II cold-slumped glass and the all-*Al* mirrors. We used a spectrophotometer of type CM-2500d [17], covering the wavelength range from 360 to 740 nm with a resolution of 10 nm (specification by the manufacturer[3]), with a maximum variation of the measured specular reflectivity depending on the measurement position of 1%. We measured the reflectivity on twelve different positions across the mirror surface, four at the mirror edges, four at the corners and four at the center, every time testing a small area of few mm². We found that the coating layer thickness of the glass mirrors produced via physical vapor deposition [8] was relatively homogeneous (see Figure 6).

---
[3] Konica Minolta Sensing, Inc., 3-91 Daisennishimachi, Sakai-ku,Sakai, Osaka, Japan.

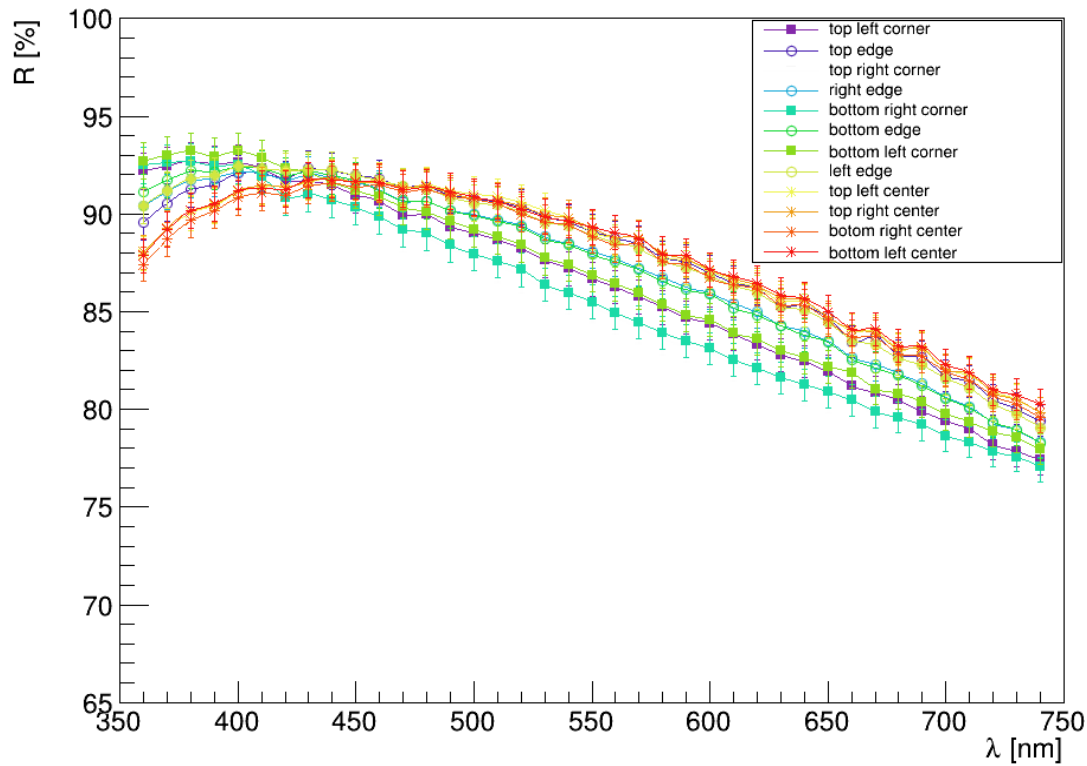

**Figure 6:** Spectral surface reflectivity of a quartz-coated and aluminized glass mirror measured at 12 different positions.

The all-*Al* mirrors were coated via plasma enhanced chemical vapor deposition. The coating is achieved by the so-called plasma polymerization. Plasma polymerization is a radical-initiated polymerization procedure. Usually, an operational gas, called precursor, is activated and fragmented by a plasma, by what radicals are created. These fragments are subsequently deposited through polymerization on the surface that needs to be coated (e.g. [18]).

Given the use of a framework in which the all-*Al* mirrors were installed while being coated, the gas underwent turbulent flow due to the narrow gap between the edges of the mirrors and the framework (private communication from Gabriele Neese at the Fraunhofer Institute[4]). We found the coating to vary in thickness, being thicker close to the mirror edges. This shifts the maximum reflectivity towards longer wavelengths (see Figure 7). Consequently, the reflectivity is significantly reduced in the wavelength range between ~360 and 400 nm, where the Cherenkov light intensity is very high.

---

[4] Fraunhofer Institute for Manufacturing Technology and Advanced Materials (IFAM), Wiener Str. 12 D-28359 Bremen, Germany [19].

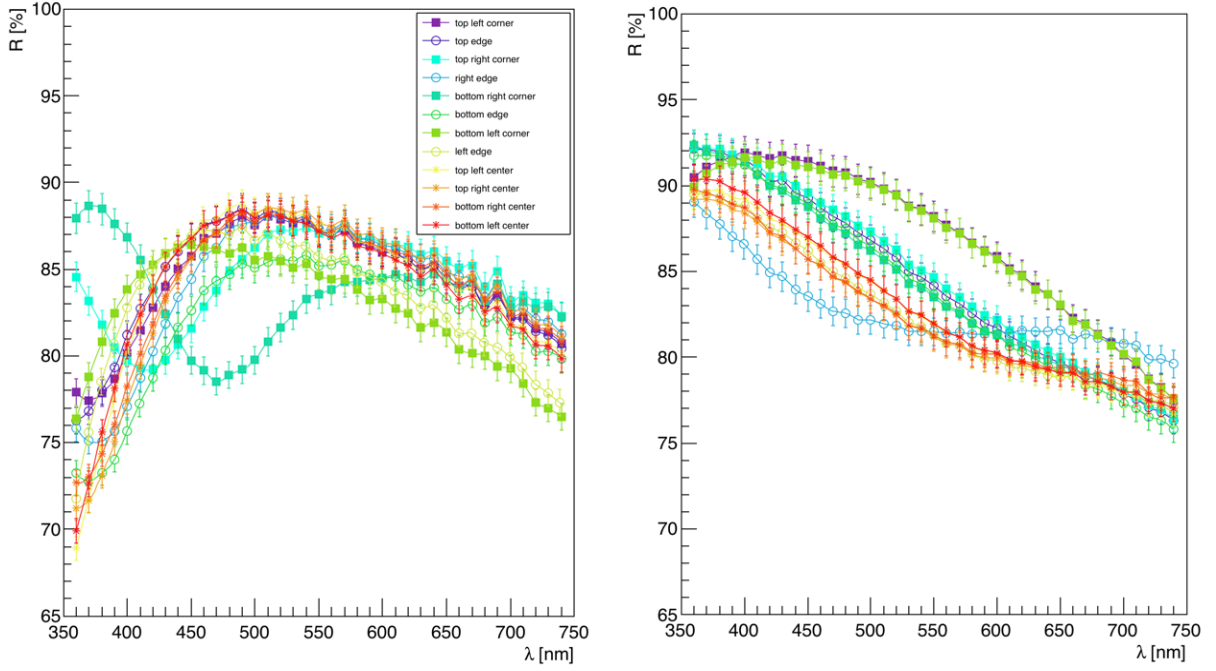

**Figure 7:** Spectral surface reflectivity of two all-*Al* mirrors measured at twelve different positions. *Left:* Inhomogeneously coated all-*Al* mirror. *Right:* More homogeneously coated all-*Al* mirror.

## III. Details of the refined experimental technique for measuring the absolute reflectance

In this section, we will present an application of the method from [12] for measuring the absolute reflectance of the MAGIC IACTs. The method was applied for some measurements similar to those presented in [12], which have been carried out in the framework of [20]. We will first describe the basic setup and the measurement technique before describing the data analysis and presenting the measurement results.

### 1. Description of the technique

The method is based on the use of a high-quality, large dynamic range CCD camera, fixed near the center of the mirror dish, which in a single shot takes the picture of a bright star and its image in the focal plane, produced by the reflector of the telescope [12]. Light coming from the star is focused by the reflector into the focal plane where the diffusely reflecting pop-up target is placed. Note that the CCD camera will capture only a tiny amount of the diffusely reflected light. An example of a CCD image taken with the described setup for MAGIC II is shown Figure 8.

In an ideal case, the absolute reflectance $R_{foc}$ of the reflector can be measured using the following equation from [12]:

$$R_{foc} = \frac{\phi_{reflection}}{\phi_{star}} \cdot \frac{\Omega_{eff} \cdot r^2}{A_{reflector} \cdot R_{target}} \quad (3)$$

where $\phi_{star}$ and $\phi_{reflection}$ denote the light flux measured from the star and is the reflected light flux imaged on the target, respectively. Both values are obtained from the sections

(white and black squares) in the CCD image shown in Figure 8.

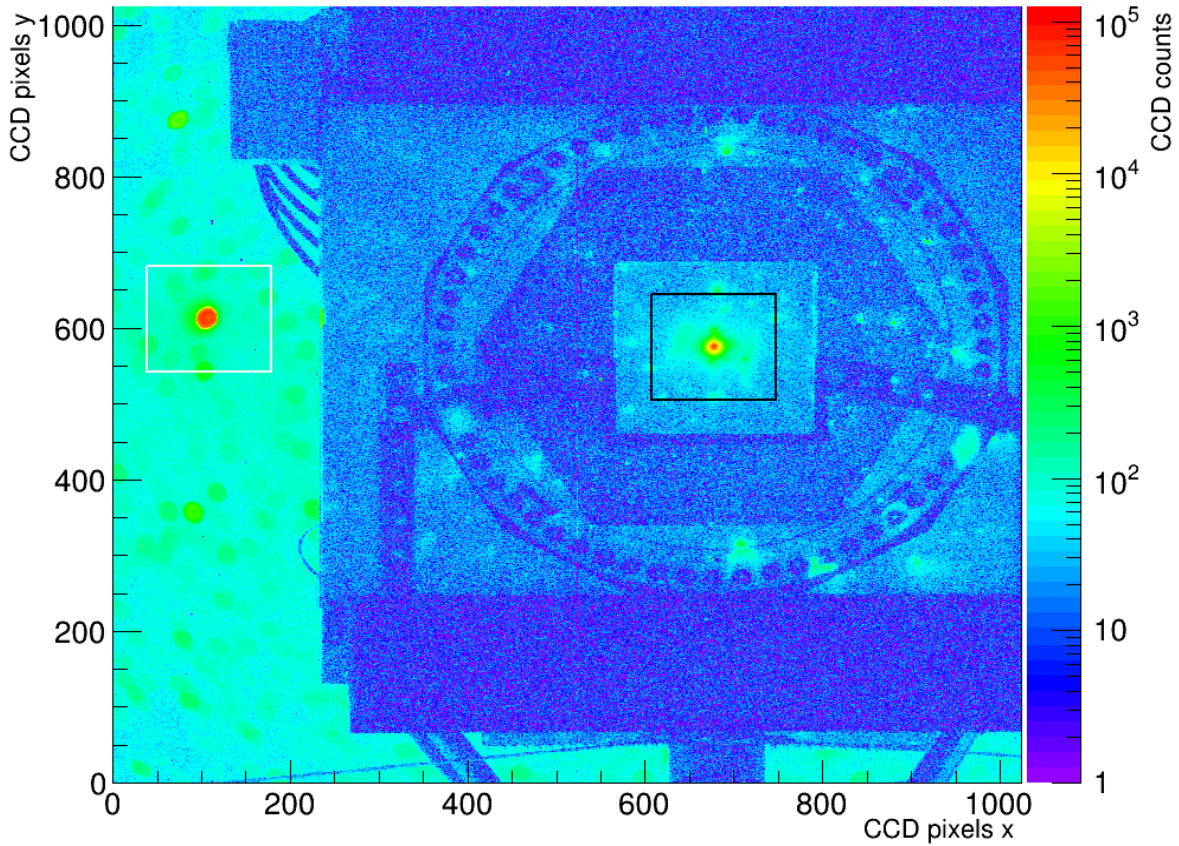

**Figure 8:** An image recorded with the CCD camera mounted near the center of the MAGIC II mirror dish. The image of the selected star (within white square) is slightly defocused. The image of the same star (black square) produced by the telescope reflector is projected onto the surface of the square-shape diffusely reflecting pop-up target.

$A_{reflector}$ is the effective area of the reflector, taking the shadowing due to the IACT camera, the supporting masts and the steel cables attached to it into account. $R_{target}$ is the reflectivity of the target, and $r$ is the distance from the target to the CCD camera lens. In order to derive the total amount of diffusely reflected light, we need to calculate the effective solid angle $\Omega_{eff}$ under which the light is reflected from the target. From the normalized angular distribution $f_{(\varphi,\theta)}$ we calculate $\Omega_{eff}$:

$$\Omega_{eff} = \int_{\theta_1}^{\theta_2} \int_{\varphi_1}^{\varphi_2} f_{(\varphi,\theta)} \cdot \sin\varphi \cdot d\varphi \cdot d\theta \tag{4}$$

where $f_{(\varphi,\theta)}$ is the reflected light distribution (for details see [20]). For a perfect diffuse reflector with a cosine angular distribution this value is equal to $\pi$ steradians.

In the following part, we will describe the measurement of the absolute reflectance. We will first describe the used equipment and later show the data reduction.

### a) Selection of a star

Ideally, a star should fulfill the following requirements to be selected as a point-like source of light:
1. High apparent brightness.
2. Single star (no double system).
3. Culmination close to local zenith.

A short exposure time for a bright star provides a high signal-to-noise ratio. Choosing a single star simplifies the data processing. Furthermore, images taken close to zenith are less affected by the atmospheric attenuation of light as well as provide low level of the Light of Night Sky (LoNS).

### b) The CCD camera

We are using a temperature-controlled CCD camera of type SBIG STL 1001E [21], equipped with a Nikon *AF NIKKOR* lens of 180 mm focal length. For fixing the aperture and the distance to the target, we set an external custom-made 40 mm diameter diaphragm in front of the lens. A set of blue, green, red, luminance and clear filters are fixed on a remotely controlled rotating wheel, allowing the measuring of the absolute reflectance for different wavelength bands.

### c) The diffuse reflecting target

We are using a target made of Spectralon as a diffuse reflector in the focal plane, which can be remotely placed right in front of the camera. This material consists of Fluor-Carbohydrate molecules, where the extremely high electronegativity of fluorine results in a very strong bond and shows extreme resistance to acids and bases. Even after several years of tests in which the samples were immersed into sea salt water and exposed to the UV light from the sun they showed almost no degradation of the reflectivity [22]. This material is supposed to show a nearly perfect Lambertian behavior as expected from an ideal diffuse reflector.

Since the knowledge of the radiation characteristics is important for the precise measurement of the absolute reflectance, we have studied the reflectivity of several samples of custom-made and industrial Spectralon [7], [20]. Our measurements showed a non-negligible deviation from the Lambert's law, finding a somewhat higher reflectivity in the forward direction (see Figure 9). An ideal diffusely scattering (Lambertian) surface scatters light according to the cosine law. For an ideal Lambertian scattering material with maximum amplitude of 1 the integral over $2\pi$ steradian (half-sphere centered on the normal direction) is $\pi$. For easier comparison we therefore normalized our measurements such that their integrals are equal to that of the cosine function. Our plot shows the average fit result of the measurements of the diffusely scattered light intensity versus the incident angle for different pieces of Spectralon.

This behavior can be understood by assuming that light penetrates to deeper layers and also these contributes to the reflection. Hence, when observing the reflected light at larger angles, due to self-absorption over longer paths, the inclined beam suffers some attenuation (our toy Monte Carlo simulation confirmed this scenario). Also, we confirmed that as long as the material was thicker than 6-7 mm, different samples of Spectralon did not exhibit any major differences in reflectivity. The measured scattering characteristic can be fitted best with:

$$f(x) \approx \cos(x)^\alpha \qquad (5)$$

where α, averaged over several Spectralon samples, was found to be $1.15 \pm 0.02$.

Higher forward reflection leads to a lower effective solid angle than $\pi$ *sr*. By taking the measured characteristic into account and by applying the Equation (4), we calculated for the wavelength 532 nm an effective solid angle $\Omega_{eff}$ of $(2.92 \pm 0.03)$ *sr* as the average of all tested samples. The total reflectance of the target was found to be $(98.6 \pm 1.2)\%$ [20].

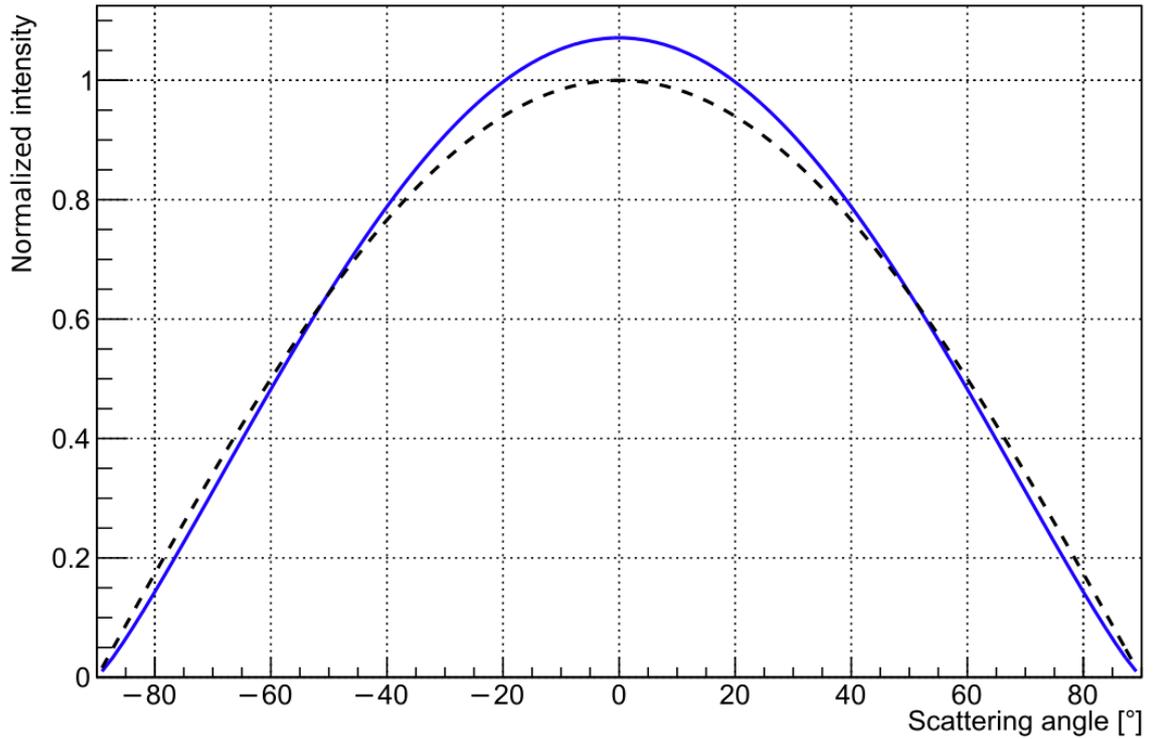

**Figure 9:** Measurement of the diffusely scattered light intensity of Spectralon samples versus the incident angle (solid curve). The integral intensity is normalized to that of cosine function (dashed curve).

We performed also some simple tests for redundancy. We illuminated with a ~3 mm diameter beam of λ = 532 nm a PIN diode of the size of 28 x 28 mm² and measured the beam intensity. Then, while illuminating a 1 cm thick piece of Spectralon with the same beam from perpendicular direction, we set the diode to its rear surface, centering the position with respect to the impinging beam. The exiting intensity that was integrated by the diode and normalized to the impinging beam, was about 1%, in good agreement with the results shown before.

We measured that the CCD camera, mounted near the reflector's center, is fixed at an angular offset of 4.24° with respect to the axis of the telescope. We used the measured angular dependence of the reflectance of the Spectralon target for correcting for the offset in Equation (3):

$$R_{foc} = \frac{\phi_{reflection}}{\phi_{star}} \cdot \frac{\Omega_{eff} \cdot r^2}{A_{reflector} \cdot R_{target}} \cdot \frac{1}{\cos(4.24°)^{1.15}} \quad (6)$$

and calculated the absolute reflectance values shown in this report.

The light fluxes $\phi_{reflection}$ and $\phi_{star}$ in Equation (6) can be extracted from a single CCD image. In the following we describe in detail the required measurement and analysis steps.

### 2. Absolute reflectance of the reflectors of the MAGIC telescopes

For our study, we acquired pictures like the one shown in Figure 8. We took images of several stars, each with a number of different filters. The CCD chips of the cameras were cooled down to -10°C during data taking to reduce the thermal noise. We used the AMC system of both telescopes to focus images of stars onto the front surface of the diffusely reflecting target. In all of these measurements the CCD camera lenses were focused to provide sharp images from the front surface of the diffuse targets. Consequently, the images of stars from infinity were somewhat blurred but this was rather an advantage; sharing the intense starlight over many pixels extends the dynamic range of the camera.

Along with the measurements of the absolute reflectance of the MAGIC-I and the MAGIC-II telescopes, we performed measurements to compare the reflectance of the all-*Al* and the glass mirror types of the MAGIC II telescope. Below we will explain in detail the data analysis.

### a) Data analysis

To measure the absolute reflectance, we need to calculate $\phi_{star}$ and $\phi_{reflection}$ from the recorded exposures (e.g. Figure 8). The data analysis chain consists of 5 steps:
- Dark frame subtraction
- Selection of the two regions of interest in the image
- Correction for the different background level in the individual sections
- Measure of $\phi_{star}$ and $\phi_{reflection}$
- Measure of $R_{foc}$ by successively increasing the integration range of $\phi_{reflection}$

#### (i) Dark frame subtraction

In the first step, we subtract from every given pixel of an exposed image its content integrated during the dark frame of the same duration. In Figure 10 we show the dark frame subtracted picture taken with the MAGIC-II CCD camera centered on the sections around the star and its reflected image on the Spectralon.

#### (ii) Selection of the two regions of interest

Image sections of 140 x 140 CCD pixels are selected around the star and its reflected spot on the diffusely reflecting target (see Figure 10, see also Figure 8). To obtain comparable results, we used the same size-cropped image sections for both telescopes. From these sections, we calculate the $\phi_{reflection}$ and $\phi_{star}$.

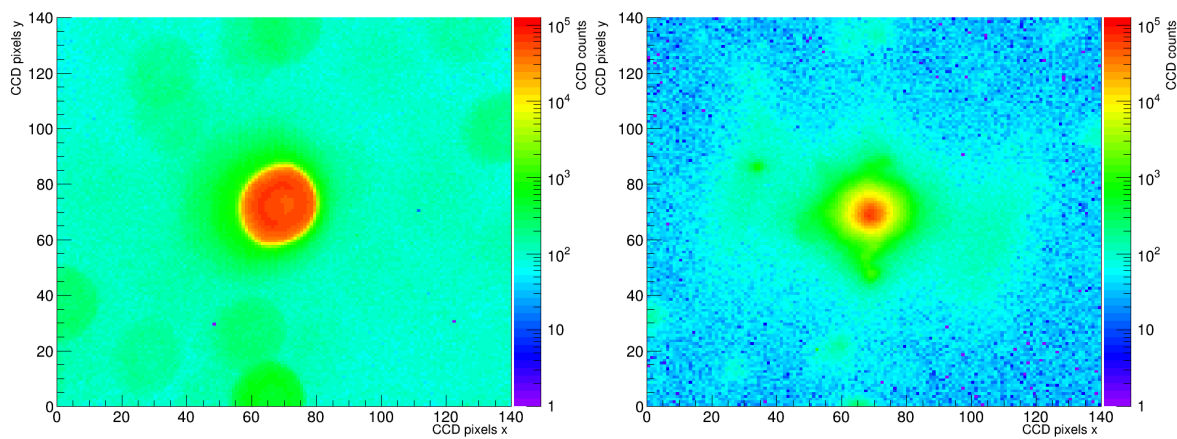

**Figure 10:**  Dark-frame-subtracted CCD frame of a star (left) and its reflected image on the diffusely reflecting target in front of the MAGIC II imaging camera (right).

#### (iii) Correction of the different background levels around the star and its reflected image

The content of individual CCD pixels in the background region of the star image (Figure 10 left) is higher than in the corresponding region in the reflected image section (Figure 10 right). We subtract from the regions selected around the star and its reflection their individual backgrounds. The procedure described below is performed separately for the two sections. To subtract the background, we use the histogram of the frequency distribution of the CCD pixel content (see Figure 11). One expects most pixels around the star will detect a low amplitude noise from the LoNS. Since we are interested in subtracting the background, in Figure 11 we zoom in to the part of the distribution containing the low amplitude counts due to the LoNS.

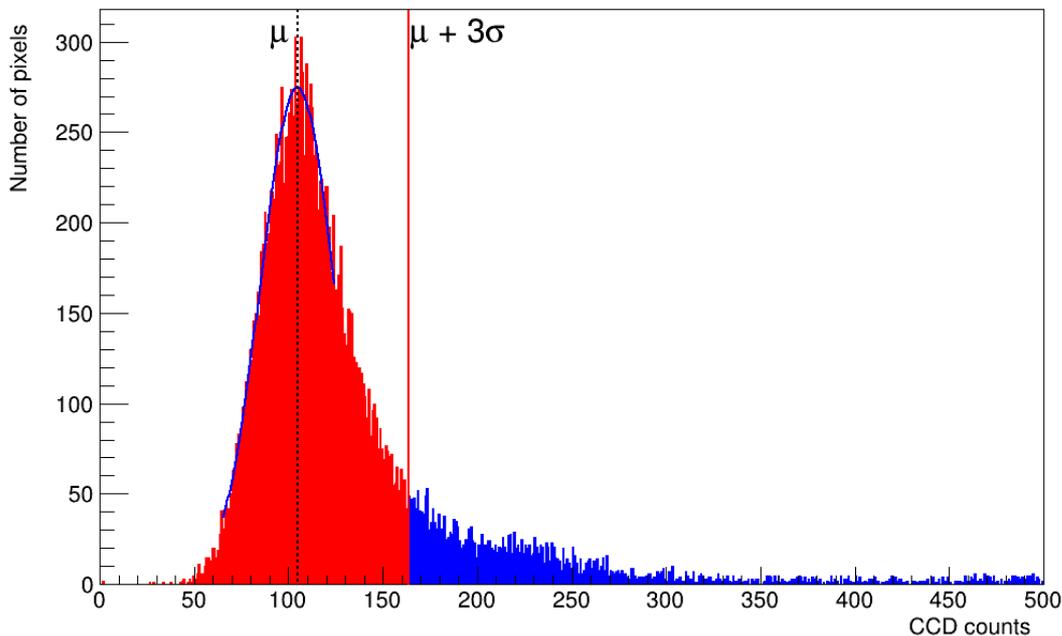

**Figure 11:** Frequency distribution of CCD counts in the star region after dark fame subtraction. The vertical solid red line denotes the three standard deviations from the mean and the vertical dashed line denotes the mean.

The noise due to LoNS shows a peak in the frequency distribution at about 100 CCD counts. As a first step, we fit a Gaussian function to the noise peak and measure its mean µ and its standard deviation σ.

We checked the following condition for each pixel in the squared section to distinguish between background and signal:

| background | $Q_i - \mu \leq 3\sigma$ | will be set to | $Q_{i,c} = 0$ |
| signal | $Q_i - \mu > 3\sigma$ | will be set to | $Q_{i,c} = Q_i - \mu$ |

where $Q_i$ is the count of $i_{th}$ pixel and $Q_{i,c}$ denotes the background subtracted charge content of the same pixel. In Figure 12 we present the sections around the star and its reflection on the target after such cleaning. Apparently, the star image contains some remnants from other stars neighboring stars. As long as these do not intersect with the region of interest, they will not affect the analysis. The laterally extended tails on the reflected image are the all-*Al* mirror-specific aberrations discussed in the precious sections. The individual, small but bright "islands" in the region around the reflected star image are due to a few mirrors.

### (i) Calculation of $\phi_{star}$

For measuring $\phi_{star}$, we sum up the CCD counts in the defined squared region of the star section. Due to the complex shape of the superimposed spots of all all-*Al* mirrors, the measurement of $\phi_{reflection}$ in the reflected section is somewhat more complex.

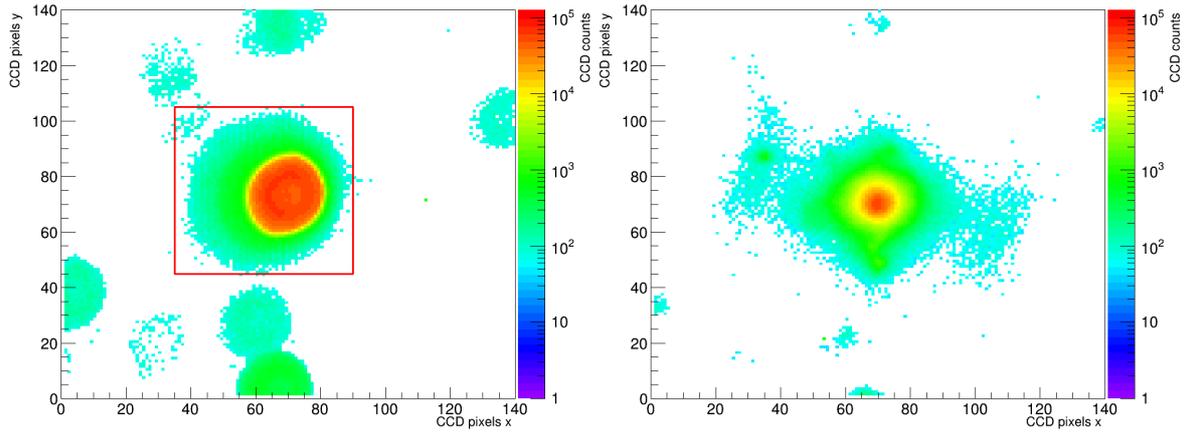

**Figure 12:** Background-corrected image around the star region (left) and the reflected image of the same star produced by the telescope's mirror dish (right) after the image cleaning.

*(ii) Calculation of $R_{foc}$*

At first, we determine the center of mass of the reflected star image. Then we calculate the amount of reflected light in a circle of a smallest radius around the center of the reflected light distribution. By using Equation (6) we derive the absolute reflectance. Then we increase the integration radius of concentric circles by one CCD pixel at a time, repeating after each increment the calculation. The results are shown in Figure 13.

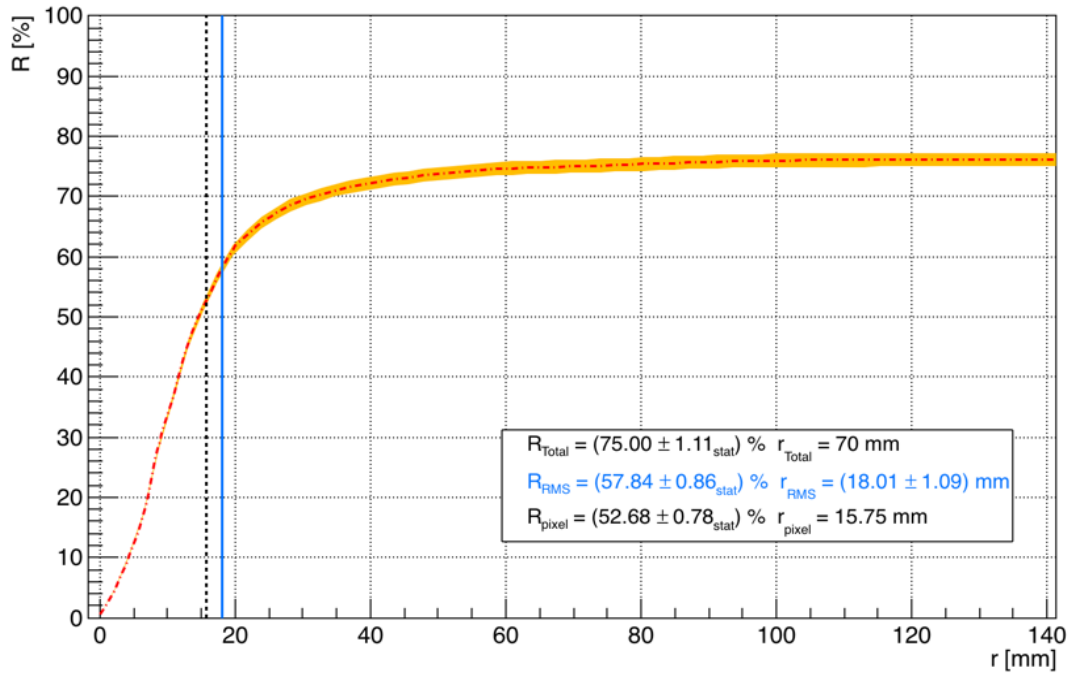

**Figure 13:** Example of the dependency of the absolute reflectance (dotted-dashed line) on the integration radius ($R_{pixel}$: vertical dashed line, $R_{RMS}$: vertical solid line) centered on the reflected star image. The shaded band around the dotted-dashed line indicates the measurement uncertainty.

We calculate the fraction of light $R_{RMS}$ that passes through the calculated RMS radius $r_{RMS}$ and consider it to correspond to the absolute reflectance of the reflector. $r_{RMS}$ corresponds to the root-mean-square distance of the CCD counts $N_i$ from the center of mass of the star image on the Spectralon target:

$$r_{RMS} = \sqrt{\frac{\sum_i N_i r_i^2}{\sum_i N_i}} \quad , \quad R_{RMS} = R(r_{RMS}) \qquad (7), (8)$$

This definition has been chosen, since it is less dependent on the actual shape of the PSF spot. We limit ourselves for the RMS calculation to a region with a radius of ≤70 mm around the center of mass of the reflected star image. This choice allows us to largely suppress the possible contribution from the reflection of nearby stars and/or accidentally defocused mirrors on the diffusely reflecting target. We furthermore calculate the absolute fraction of light $R_{pixel}$ that is concentrated in the equivalent aperture of a single pixel of the MAGIC imaging camera. The latter has a hexagonal input window with an inner radius of 0.05°. Another relevant number is the reflectance $R_{Total}$ within a circle of 140 mm diameter (note that this is equivalent to an angular diameter of almost 0.5°). As shown in Figure 13, the absolute reflectance continues to increase up to a radius of 70 mm, where it almost saturates. The following slow increase of the quasi-saturated curve indicates that there will be more light outside of the chosen integration radius of 70 mm. The directional-diffuse component of light around the direction of specular reflection can also contribute into this. Obviously, the large spot size of 140 mm diameter is too big for any meaningful gamma-hadron separation (see Section I). For that very reason it makes no sense to refer to this as the mirror reflectance of a Cherenkov telescope.

By using band-pass filters, we measured the absolute reflectance for several wavelengths for both MAGIC telescopes according to the method described above. Here we present the mean values over the measurements performed with the blue filter (see Table 1). As our active mirror control allows us to focus each mirror individually, we were able to perform measurements grouping the two mirror types, that are installed on the MAGIC II dish, separately.

Table 1: Mean results of the absolute reflectance measured with the blue filter for both MAGIC telescopes. In the case of MAGIC II, the measurement was repeated separately for the different mirror types.

| Telescope/Mirror type | $r_{RMS}$ [mm] | $R_{RMS}$ [%] | $R_{Pixel}$ [%] | $R_{Total}$ [%] |
|---|---|---|---|---|
| **MAGIC I** | 19.3 ± 1.1 | 55.0 ± stat 0.5 sys. 1.0 | 48.8 ± stat. 0.5 sys. 0.9 | 71.2 ± stat. 0.7 sys. 1.3 |
| **MAGIC II** | 16.1 ± 1.1 | 55.1 ± stat. 0.5 sys. 1.0 | 54.5 ± stat. 0.5 sys. 1.0 | 74.7 ± stat. 0.6 sys. 1.3 |
| **MAGIC II all-*Al*** | 14.4 ± 1.1 | 54.2 ± stat. 0.7 sys. 1.0 | 55.9 ± stat. 0.7 sys. 1.1 | 67.2 ± stat. 0.9 sys. 1.3 |
| **MAGIC II glass** | 16.2 ± 1.1 | 56.3 ± stat. 0.6 sys. 1.0 | 54.9 ± stat. 0.6 sys. 0.9 | 81.4 ± stat. 0.9 sys. 1.4 |

Comparing the $r_{RMS}$ of the focused spots of the two telescopes, one can see that the MAGIC-II mirrors can be better focused than those of MAGIC I. Also, the all-*Al* type mirrors of MAGIC-II provide better focusing than the glass mirrors. Our measurements show an average $R_{pixel}$ of $(48.8 \pm 0.5_{stat} \pm 0.9_{sys})\%$ for MAGIC-I and $(54.5 \pm 0.5_{stat} \pm 1.0_{sys})$ for MAGIC-II. As mentioned above, $R_{Total}$ is only of academic interest. It is interesting to note that compared to the all-*Al* mirrors, the glass mirrors on the MAGIC-II telescope show a significantly higher value for the $R_{Total}$, due to much lower diffuse light scattering.

In order to verify the robustness of this measurement method, we repeated the measurement on selected stars multiple times on different nights and under different zenith angles. While $R_{Total}$ was well reproduced, the values of $r_{RMS}$, $R_{RMS}$ and $R_{pixel}$ showed some variation. This can be explained by the different temperatures and hence the emission spectra of selected stars as well as by the variations of the mirrors' surface reflectivity with wavelength. Also, the non-negligible transmission of filters outside their defined transmission bands as well as the small variations related to the AMC system performance at varying observational angles may influence these results.

### b) Error estimation

To estimate the error of the absolute reflectance measurements, we took various uncertainties into account, most of which are of systematic nature. The uncertainty of the effective mirror area, due to shadow cast by the masts holding the imaging camera of MAGIC, the steel cables for reinforcing the camera holding structure and the imaging camera by itself was estimated to be 1.5 m². Another source of error is the distance between the mirror front surface and the Spectralon target (< 0.05 m). The uncertainties of the effective solid angle (0.03 sr) and of the total reflectance of the Spectralon target (1.2%) are estimated from the variation of the measured values for different samples [20]. All above-mentioned uncertainties were combined into a systematic uncertainty using Gaussian error propagation.

We applied both the upper and lower limits of the error of the Gaussian fits to the background histograms for estimating the uncertainty of the image cleaning procedure for measuring the $\phi_{star}$ and $\phi_{reflection}$. Limits for the light content in the star and its reflected image are calculated for both signs of the error and the mean of the two peak values is used as uncertainty for the light content in those regions (for more detail on the error calculation we refer to [20]). Those uncertainties are of statistical nature and vary from measurement to measurement. We used them to compute a weighted average over different measurements on different stars and state the error of the weighted average as statistical uncertainty. We neglect the uncertainty of the correction for the small offset of the SBIG-camera with respect to the optical axis. For $r_{RMS}$ we state an error that corresponds to half the size of a CCD pixel.

### 3. Discussion

We have further refined and applied *in situ* the method [12] for measuring the absolute reflectance of the concave mirrors for the MAGIC-I and MAGIC-II telescopes. Especially we performed in-depth studies of the scattered light component of the mirrors. The obtained results are somewhat unexpected and surprising. While $r_{RMS}$ is in the range of ~15 - 20 mm, the absolute amount of light concentrated within these radii is rather low. The absolute amount of integrated light within a diameter of ~0.5° (140 mm) is ~70 %. Such radius is too big for any meaningful gamma-hadron separation; we mention this value just as a matter of fact. Only about 50 % of impinging light is contained within a single central pixel of 0.10° (30 mm) in diameter of the MAGIC telescopes imaging cameras. The remaining ~20 % of light is distributed within a ring segment of ~(0.10 - 0.5)° in diameter. By assuming that an average good mirror can reflect about 80 - 85% of light, it becomes obvious that ≥ 10% of light is scattered beyond the central area of ~0.5° radius. An indirect evidence of this can be seen in Figure 13; while the cumulative amount of concentrated light should saturate for a relatively large integration radius, in fact it still keeps increasing, albeit at a lower rate.

Results for off-axis spots, exclusively used in the IACT imaging technique, should be even worse than for the central pixel.

Local deviations of the mirror surface curvature from its designed value can be considered as one main reason for larger than anticipated losses. Such deviations may have their origin in the production process but also with time and ageing the "frozen" tensions in the composite, light-weight glued mirrors, together with possible tension due to the mechanical fixation and adjustment mechanism on the reflector, may show up. This may affect not only the diamond-milled all-*Al* mirrors but also the composite glass mirrors. With ageing larger deviations of the radius of curvature could be anticipated, especially at the edges of the mirrors.

We found out with time some number of all-*Al* mirrors on MAGIC-I got kind of wavy structures, "bulges", on their surface and even in a couple of cases the front aluminum plate of the small mirrors got disintegrated. We think that water and-humidity from rain and clouds had a chance to penetrate through microscopic holes in the sealing of some mirrors, accumulating to macroscopic amounts of water inside their volume. It is easy to imagine that in some cases those water "clusters" freeze during minus temperatures and as a result of expansion change locally the mirror's radius of curvature. Obviously, such mirrors collect less light in the focal plane.

A plausible explanation for the steady increase of the measured amount of light with increasing radius (see Figure 13) can be the fact that there is more diffusely scattered light than what was assumed.

Here it is important to recall that one differentiates between three types of reflectivity: specular, directional diffuse and diffuse. The large surface roughness of up to ~13 nm for the all-*Al* type is the reason for the diffuse and directional diffuse components. The directional diffuse reflection component can have a contribution into the big halo around the central spot (see Figure 12, right).

That a significant part of light cannot be focused within one single pixel is important especially at the threshold of IACTs when the scarce light from the low energy air showers becomes distributed over a large number of pixels, producing a kind-of optical cross-talk. This reduces the genuine signals of individual channels and sets a higher threshold for the trigger and the subsequent analysis.

In [23] the authors discuss how scattering due to micro-roughness theoretically affects the PSF and thus the so-called encircled energy of mirrors used by IACTs. Unfortunately, no direct comparison with our measurements of the absolute reflectance can be made for different reasons. First, the normalization procedures are different. While the measurement presented here is based on an absolute calibration w.r.t. the measured light flux of a star incident to the IACT reflector, including also the reflectivity of the mirror surface material, Tayabaly et al. [23] do not include the latter in their calculations, considering only the surface geometry of the mirror. In addition, the authors assume an incident power of one, thus putting no limiting assumptions on the size of the detector and therefore accounting for the whole scattering. By contrast, the absolute reflectance measurements we present here were only measured up to a certain radius (see Figure 13). Finally, for our measurements, a blue bandpass filter was centered on 450 nm was used, while Tayabaly et al. [23] present encircled energy profiles that were calculated for 300 and 550 nm. Since both, the surface reflectivity and the scattering due to micro-roughness are wavelength-dependent, it is thus not straightforward to perform a detailed comparison between these results.

An interesting observation can be made concerning the longevity of the all-*Al* type mirrors: in spite of their initially relatively low absolute reflectance, their surface reflectivity showed no degradation during a time span of three years. The reason is most likely the few mm thick, massive *Al* plate in the front. Unlike the glass mirrors that typically are covered by a ~100 nm thick layer of *Al* below the protective layer, it cannot be easily attacked by the impinging flux of aggressive acid components contained, for example, in rain water during lightning. The latter can find its way through the micro-cracks in the protective quartz layer of the mirrors, originated during the coating process (private communication from G. Pareschi from INAF) as well as these micro-cracks can be produced due to warm/cold cycles and the "bombardment" by sand brought by wind in desert and mountain locations. Unlike the all-*Al* mirrors, the liquid penetrated through micro-cracks in the glass mirrors cannot attack the glass and thus it can spread only in the transverse direction, attacking the thin *Al* layer. With time and many cycles of warm-up and cooling down, large areas in the thin *Al* layer can become oxidized or change their chemical composition, thus seriously reducing the mirror reflectivity.

## IV. Conclusions

We significantly refined and improved the precision of the absolute reflectance measurement method, originally described in reference [12]. Especially the light scattering effects by mirrors were studied in great detail. Furthermore, we developed a simple approach for measuring the amount of scattered light due to dust deposit and scratches on the mirrors. We measured that Spectralon, the best known diffuse reflecting material, deviates from the ideal Lambertian scatterer, reflecting more light in the forward direction. The *in-situ* method we present here can be applied to any telescope and can provide an easy and precise method for monitoring the absolute reflectance.

This study showed that although the chosen optical design of the MAGIC telescope and its mirrors show a satisfactory performance for the used pixel size, these can be significantly improved by using production technologies that can guarantee low surface roughness of $\leq 2$ nm, a protective layer of constant thickness and a more accurate and stiff geometry of the mirror surface.

In addition, it is important to avoid that the mirror fixation mechanism on the reflector of a telescope transmits any tension to the mirror, which otherwise could deform its shape.

Note that typically an IACT performs measurements during ~10% of time; in the remaining 90% of time its mirrors are exposed to weathering, usually in harsh mountainous and desert locations. Using a protective layer of significantly higher durability than the currently used ones, has the promise to provide both a low level of diffusely-scattered light from the front surface as well as long-lifetime of the reflecting layer.

The above reported systematic studies started in 2008 and were carried out over a time span of about six years. Our studies helped us to better understand the absolute reflectance and to refine the measurement technique. As a consequence, in the last three years we replaced part of the low reflectance mirrors on both the MAGIC-I and the MAGIC-II telescopes. The new mirrors of improved quality enhanced the reflectance by ~20 %. We are planning to devote more time and efforts to the mirror maintenance issue of the MAGIC telescopes.

**Acknowledgments**

We want to gratefully thank Adrian Biland from ETH Zurich for the modifications in the AMC software that helped us to perform some of the important measurements.

The authors feel indebted to their colleagues from the MAGIC collaboration for their continuous support and interest during the performance of this project.